\documentclass[%
 reprint,
 superscriptaddress,
 amsmath,amssymb,
 aps,
]{revtex4-1}
\usepackage{xcolor}
\usepackage{graphicx}
\usepackage{dcolumn}
\usepackage{bm}
\usepackage{ulem}
\begin{document}
\preprint{APS/123-QED}

\title{Control of magnon-photon coupling by spin torque}

\author{Anish Rai}\email{arai@udel.edu}
\affiliation{Department of Physics and Astronomy, University of Delaware, Newark, Delaware 19716, United States}
\author{M. Benjamin Jungfleisch}\email{mbj@udel.edu}
\affiliation{Department of Physics and Astronomy, University of Delaware, Newark, Delaware 19716, United States}

\date{\today}

\begin{abstract}
We demonstrate the influence of damping and field-like torques in the magnon-photon coupling process by classically integrating the generalized Landau-Lifshitz-Gilbert equation with RLC equation in which a phase correlation between dynamic magnetization and microwave current through combined Amp\`ere and Faraday effects are considered. We show that the gap between two hybridized modes
can be controlled in samples with damping parameter in the order of $10^{-3}$ by changing the direction of the dc current density $J$ if a certain threshold is reached. Our results suggest that an experimental realization of the proposed magnon-photon coupling control mechanism is feasible in yttrium iron garnet/Pt hybrid structures.
\end{abstract}

\maketitle

\section{\label{sec:level1}Introduction}
Coherent magnon-photon coupling in hybrid cavity-spintronics contributed to the advancement of magnon-based quantum information and technologies \cite{PhysRevLett.111.127003,PhysRevLett.113.083603,PhysRevLett.113.156401,PhysRevLett.113.156401,PhysRevLett.114.227201,doi.org/10.1038/ncomms9914,Hu_2015_v2,doi:10.1126/science.aaa3693,doi:10.1126/sciadv.1501286,10.1038/nphys3347,PhysRevLett.116.223601,PhysRevLett.118.217201,Zhang2017,PhysRevLett.121.203601,doi:10.1063/5.0020277,9706176}. The collective excitations of an electron spin system in magnetically ordered media called magnons can couple to microwave photons via dipolar interaction, demonstrating level repulsion and Rabi oscillations \cite{PhysRevLett.113.156401}. Strongly coupled magnon-photon systems have been explored to bring many exotic effects into the limelight, some of which include the manipulation of spin currents \cite{PhysRevB.94.054433}, and bidirectional microwave-to-optical transduction \cite{PhysRevB.93.174427,Mojtaba}. In addition to the coherent magnon-photon coupling, there exists an exciting domain of dissipative magnon-photon coupling where level attraction can be observed, which is characterized by a coalescence of the hybridized magnon-photon modes \cite{PhysRevLett.121.137203,PhysRevB.100.214426,PhysRevB.99.134426,PhysRevApplied.11.054023,Rao_2019,PhysRevB.100.094415,PhysRevB.105.214418}. 
\par

The theoretical framework of magnon-photon coupling is given by the following dispersion relation \cite{Bhoi2019PhotonmagnonCH} of the hybridized modes:
\par

\begin{equation}\label{Eq: dispersion}
\widetilde{\omega}_{\pm}=\frac{1}{2}\left[\left(\widetilde{\omega}_{m}+\widetilde{\omega}_{c}\right) \pm \sqrt{\left(\widetilde{\omega}_{m}-\widetilde{\omega}_{c}\right)^{2}+4 g^{2}}\right],
\end{equation}
where $\widetilde{\omega}_{m}=\omega_{m}-i \alpha \omega_{m}$ and $\widetilde{\omega}_{c}=\omega_{c}-i \beta \omega_{c}$ are the complex resonance frequencies of the magnon and photon (cavity) modes, respectively. $g$ is the coupling between the two modes. 
 $\alpha$ and $\beta$ are the intrinsic damping rates of the magnon and photon modes, respectively. The real and imaginary parts of $\widetilde{\omega}_{\pm}$ represent the dispersion shape and the {linewidth of the coupled modes}, respectively. The second term of the square root in Eq.~(\ref{Eq: dispersion}) not only gives the strength of the coupling but also reveals the nature of the coupling. Harder and co-workers \cite{PhysRevLett.121.137203} carefully introduced a coupling term based on the cavity Lenz effect to mitigate the Amp\`ere effect. However, the on-demand manipulation of the magnon-photon polariton by spin torques has not been addressed so far.
\par

\begin{figure}[b]
\centering
\includegraphics[width=0.9\columnwidth]{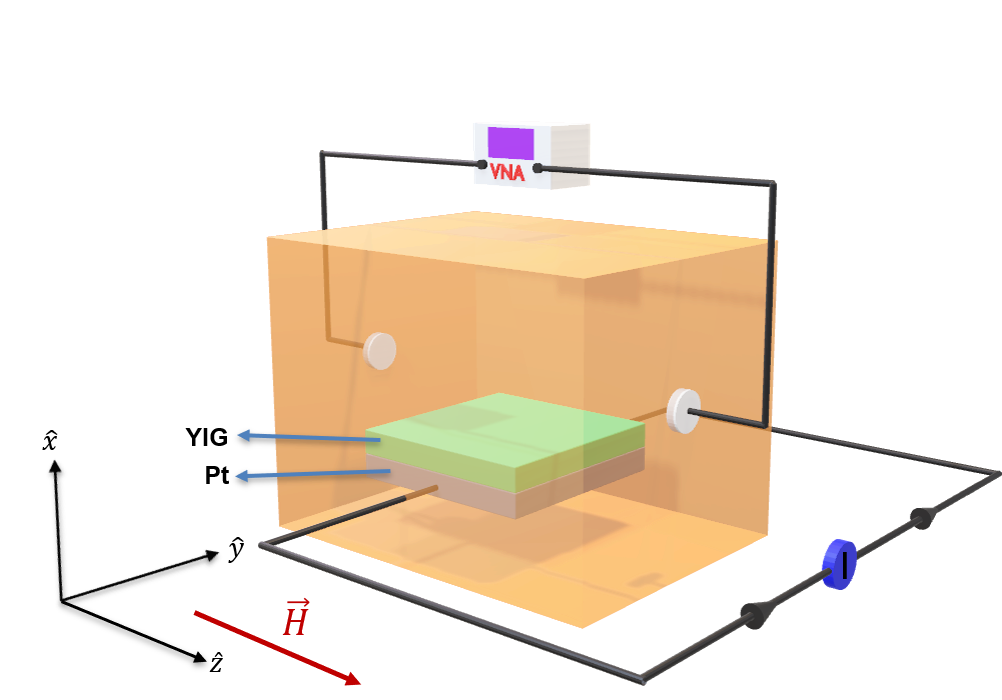}
\caption{The schematic of the experimental setup. A patterned YIG/platinum(Pt) bilayer is the sample under consideration. The dc current is passed through the platinum layer. The microwave current is passed through the cavity and analyzed using a Vector Network Analyzer (VNA). Here, the external magnetic field is applied along $\widehat{z}$ direction.} 
\label{fig:schematic}
\end{figure}

\begin{figure*}[t]
\centering
\includegraphics[height = 6 in , width = \textwidth, keepaspectratio]{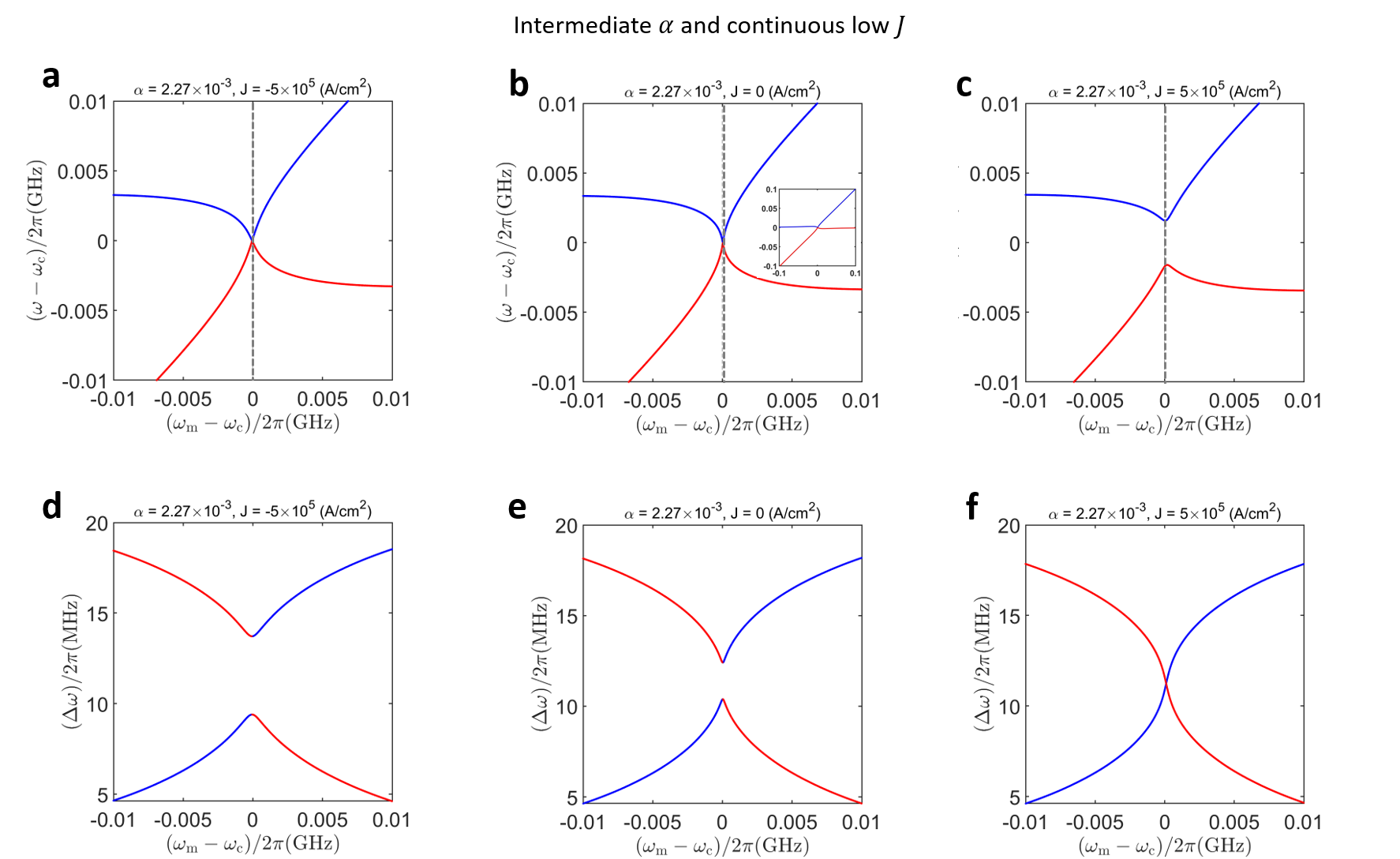}
\caption{Magnon-photon coupling control for an intermediate value of the Gilbert damping parameter $\alpha$ and a continuous, low current density $J$: The dispersion ($\omega-\omega_{c}$) in (a-c) and the linewidth ($\Delta{\omega}$)  
in (d-f) are plotted as a function of the field detuning ($\omega_{m}-\omega_{c}$) for $\alpha = 2.27\times10^{-3}$. The hybridization of magnon and photon modes is compared for different dc current densities $J$: (a), (d) J = $-5\times10^5~\mathrm{A/cm^2}$, (b), (e) J = $0~\mathrm{A/cm^2}$ and (c), (f) J =$5\times10^5~\mathrm{A/cm^2}$. The blue and red line represent the two hybridized modes. The inset in (b) shows that for larger field detuning ($\omega_{m}-\omega_{c}$), the uncoupled photon mode approaches $\omega_{c}$ making  ($\omega-\omega_{c}$) approach zero.}
\label{fig:Intermediate_alpha_low_J}
\end{figure*}

\begin{figure*}[t]
\centering
\includegraphics[height = 6 in , width = \textwidth, keepaspectratio]{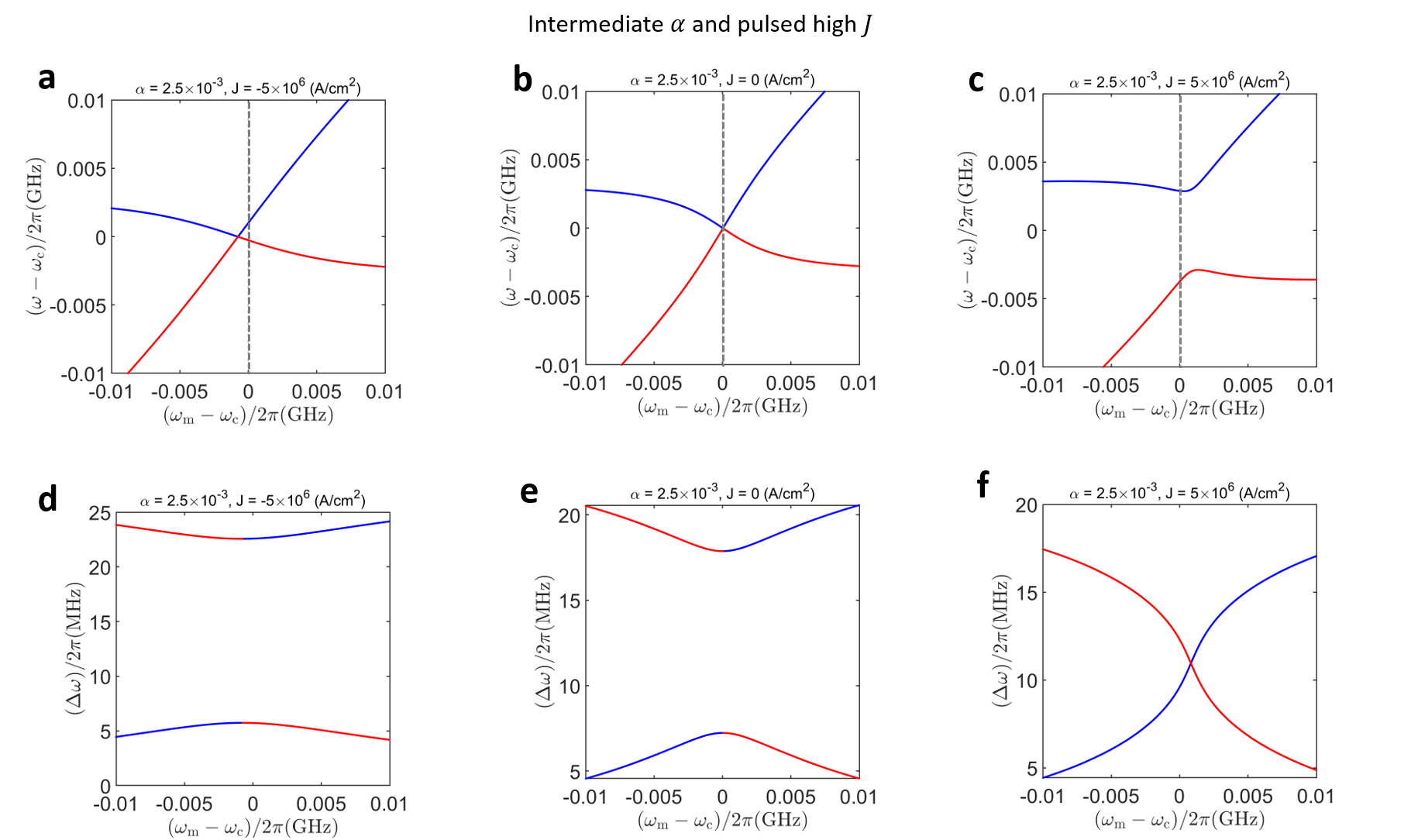}
\caption{Magnon-photon coupling control for an intermediate value of the Gilbert damping parameter $\alpha$ and a pulsed, high current density $J$: The dispersion ($\omega-\omega_{c}$) in (a-c) and the linewidth ($\Delta{\omega}$)  
in (d-f) are plotted as a function of the field detuning ($\omega_{m}-\omega_{c}$) for $\alpha = 2.5\times10^{-3}$. The hybridization of magnon and photon modes is compared for different dc current densities $J$: (a), (d) J = $-5\times10^6~\mathrm{A/cm^2}$, (b), (e) J = $0~\mathrm{A/cm^2}$ and (c), (f) J =$5\times10^6~\mathrm{A/cm^2}$. The blue and red line represent the two hybridized modes.}
\label{fig:Intermediate_alpha_high_J}
\end{figure*}

In this work, we examine the influence of damping- and field-like torques in the magnon-photon coupling process. 
Our results indicate that the magnitude of the level repulsion (manifested by the frequency gap of the hybridized modes) and, hence, the magnon-photon coupling strength can efficiently be controlled by varying the magnitude and the direction of dc current density $J$ for realistic parameters of the magnetic properties. By coupling the generalized Landau-Lifshitz Gilbert equation with the RLC equation of the cavity, we show that an on-demand manipulation of the magnon-photon coupling strength can be achieved for current densities of the order as small as $10^5$~A/cm$^2$. 
\par
This article is structured in the following fashion. In section II, we discuss the classical description to model our system, in which  the ferromagnetic resonance of the magnetic system is strongly coupled to photon resonator mode of the microwave cavity. In section III, we introduce the parameters used for the analysis followed by a detailed discussion of our findings. In section IV, we summarize our work.
\section{\label{sec:level2}Classical Description}
The magnetization dynamics in ferromagnetic systems can be described by the generalized Landau-Lifshitz-Gilbert equation \cite{landau1935theory,SLONCZEWSKI1996L1,gilbert2004phenomenological} given by:
\begin{equation}
\label{LLG}
\begin{split}
\frac{d\vec{M}}{dt}=\gamma\vec{M}\times\vec{H}_{\mathrm{eff}}-\frac{\alpha}{M_\mathrm{s}}\left(\vec{M}\times\frac{d\vec{M}}{dt}\right)+\\
\frac{\gamma a_\mathrm{J}}{M_\mathrm{s}}\left(\vec{M}\times\left(\vec{M}\times\vec{p}\right)\right)-\gamma b_\mathrm{J}\left(\vec{M}\times\vec{p}\right),
\end{split}
\end{equation}

where $\vec{M}$ is the magnetization vector, $M_\mathrm{s}$ is the saturation magnetization, $\vec{H}_{\mathrm{eff}}$ is the effective magnetic field including external field $\vec{H}$, anisotropy, microwave, and demagnetization fields, $\gamma$ is the gyromagnetic ratio, $\alpha$ is the Gilbert damping parameter, $\vec{p}$ is the spin polarization unit vector. Furthermore, the terms proportional to $a_\mathrm{J}$ and $b_\mathrm{J}$ are the damping-like torque and field-like torque, respectively. The coefficients $a_{J}$ and $b_{J}$ are defined as \cite{Pathak2020}:
\begin{equation}
\label{eta}
 a_{J} = \frac{\eta_{a} J \hbar}{2e M_sd}, b_{J} = \frac{\eta_{b} J \hbar}{2e M_sd},  
\end{equation}
where $\eta_{a}$ and $\eta_{b}$ are the damping-like torque efficiency and field-like torque efficiency, respectively. $J$ is the dc current density, whose polarity determines the directions of field-like and damping-like torque terms through Eqs.~(\ref{LLG}) and  (\ref{eta}), $\hbar$ is the reduced Planck's constant, $e$ is the electron charge, 
 and $d$ is the thickness of the ferromagnetic sample. We define the magnetic field, magnetization, and spin polarization unit vectors as $\vec{H}_{t}=h_{x}(t) \widehat{x}+h_{y}(t) \widehat{y}+H \widehat{z}$, $\vec{M}=m_{x}(t) \widehat{x}+m_{y}(t) \widehat{y}+M_{s} \widehat{z}$ and $\vec{p}=\widehat{z}$,
where $H$ and $M_\mathrm{s}$ are the dc magnetic field and saturation magnetization, respectively, and $h_{x,y} (t)$ and $m_{x,y} (t)$ are
the dynamic magnetic field and magnetization.

\begin{figure*}[t]
\centering
\includegraphics[width = 1.0\textwidth, keepaspectratio]{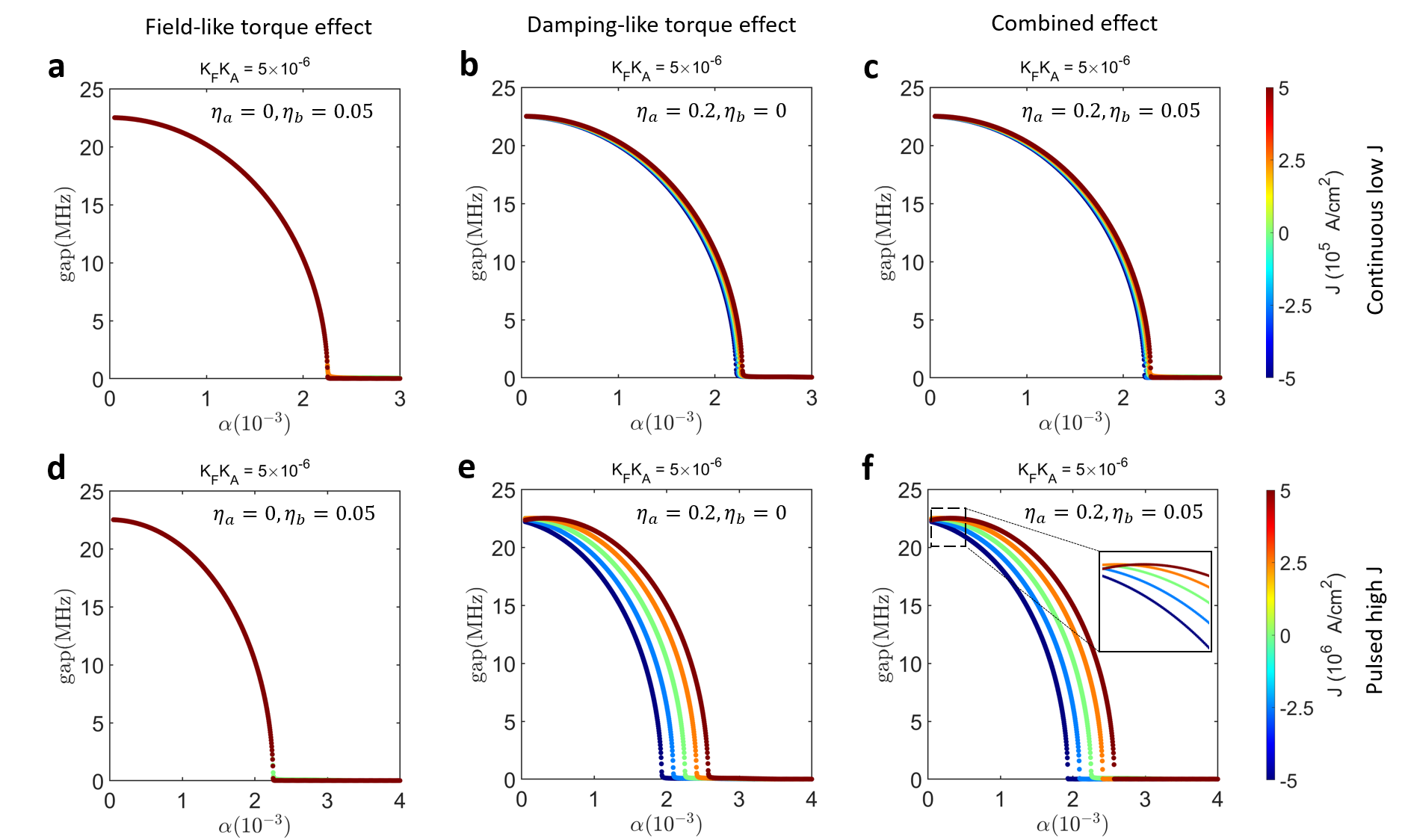}
\caption{Variation of coherent magnon-photon coupling (minimum frequency gap between two hybridized modes) for different values of $\alpha$ and $J$. For (a), (b), and (c) $\alpha$ is varied from $3\times10^{-3}$ to $5\times10^{-5}$ and $J$ (continuous) is varied from $-5\times10^5~\mathrm{A/cm^2}$ to $5\times10^5~\mathrm{A/cm^2}$ and for (d), (e), and (f) $\alpha$ is varied from $4\times10^{-3}$ to $5\times10^{-5}$ and $J$ (pulsed) is varied from $-5\times10^6~\mathrm{A/cm^2}$ to $5\times10^6~\mathrm{A/cm^2}$. Based on our model, we can distinguish between field-like contribution (a) and (d), damping-like contribution (b) and (e), and a combination of field-like and damping-like contribution to the manipulation of the anticrossing gap (c) and (f). For (a) and (d) $\eta_{a}=~0$ and $\eta_{b}=~0.05$ (pure field-like torque effect), for (b) and (e) $\eta_{a}=~0.2$ and $\eta_{b}=~0$ (pure damping-like torque effect), and for (c) and (f) $\eta_{a}=~0.2$ and $\eta_{b}=~0.05$ (combination of damping-like and field-like torque effects).}
\label{fig:variation_alpha_gap_J}
\end{figure*}

If we define the 
dynamic components, $h = h_x + ih_y$ and $m = m_x + im_y$, then 
Eq.~(\ref{LLG}) can be reduced to:
\begin{equation}
\left(\omega-\tilde{\omega}_{m}+\gamma \tilde{c}_\mathrm{J}\right) m+\omega_{s} h=0,
\end{equation}
{where $\tilde{\omega}_{m}$ is the complex ferromagnetic resonance frequency defined by $\tilde{\omega}_{m}~=~\omega_{m}-i\alpha\omega$ (where $\omega_{m}\simeq\gamma H$ is the ferromagnetic resonance frequency)}, $\omega_{s}=\gamma M_{\mathrm{s}}$, and $\tilde{c}_\mathrm{J}=b_\mathrm{J}-ia_\mathrm{J}$.
The effective RLC circuit for the cavity can be written as \cite{PhysRevLett.121.137203}:
\begin{equation}
\begin{aligned}
R j_{x,y}(t)+\frac{1}{C} \int j_{x,y}(t) d t+L \frac{d j_{x,y}(t)}{d t}=V_{0 x,y}(t),
\end{aligned}
\end{equation}
where R, L, and C represent the resistance, inductance, and capacitance, respectively. $V_{0x,y}$ is the voltage that drives the microwave current. For 
$j = j_x + ij_y$ and $V_0 = V_{0x} + iV_{0y}$, we have \cite{PhysRevLett.121.137203}
\begin{equation}
\left(\omega^{2}-\omega_{c}^{2}+i 2 \omega \omega_{c} \beta\right) j=i \frac{\omega}{L} V_{0},
\end{equation}
where $\omega_{c}=1/\sqrt{LC}$ is the cavity resonance frequency and $\beta=(R/2)\sqrt{C/L}$ is the intrinsic damping of the cavity-photon mode.\par
The microwave magnetic field will exert a torque on the magnetization through Amp\`ere's law. The relation can be expressed as:
\begin{equation}
h_{x}=K_{A}j_{y}, h_{y}=-K_{A}j_{x},
\end{equation}
where $K_{A}$ is the positive coupling term associated with a phase relation between $j_{x,y}$ and $h_{x,y}$. In a similar way, the precessional magnetization will induce a voltage in the RLC circuit through Faraday induction:
\begin{equation}
V_{x}=-K_{F}L\frac{dm_{y}}{dt}, V_{y}=K_{F}L\frac{dm_{x}}{dt},
\end{equation}
where $K_{F}$ is the positive coupling  term associated with a phase relation between $V_{x,y}$ and $m_{x,y}$. Combining Eqs.~(4)-(8) gives us the coupled equations of the form:
\begin{equation}
\begin{split}
\left(\begin{array}{cc}
\omega^{2}-\omega_{c}^{2}+i 2 \beta \omega_{c} \omega & i \omega^{2} K_{F} \\
-i \omega_{s}K_{A} & \omega-\tilde{\omega}_{m}+\gamma \tilde{c}_\mathrm{J}
\end{array}\right)\left(\begin{array}{c}
j \\
m
\end{array}\right)\\
=\left(\begin{array}{c}
i \omega \omega_{c} j_{0} \\
0
\end{array}\right),
\end{split}
\end{equation}
where $j_{0}=V_{0}\sqrt{C/L}$. The hybridized eigenmodes are calculated by solving the determinant of Eq.~(9). This yields the following analytical form [see Supplemental Material (SM)]:
\begin{equation}
 \tilde{\omega}_{\pm}=\frac{\left(\frac{\omega_{c}}{1+i \beta}+\frac{\omega_{m}-\delta}{1+i \alpha}\right) \pm \sqrt{\left(\frac{\omega_{c}}{1+i \beta}-\frac{\omega_{m}-\delta}{1+i \alpha}\right)^{2}+\frac{2 \omega_{c} \omega_{s} K_{F} K_{A}}{(1+i\alpha)(1+i\beta)}}}{2} ,  
 \label{main equation}
\end{equation}
where $\delta=\gamma\tilde{c}_{J}$. Here, $\gamma$ is the gyromagnetic ratio and $\tilde{c}_{J}$ is a complex term associated with $b_\mathrm{J}$ and $a_\mathrm{J}$ defined by $\tilde{c}_\mathrm{J}=b_\mathrm{J}-ia_\mathrm{J}$.

\section{\label{sec:level3}Results and Discussion}

For our model we choose the following realistic parameters \cite{Harder_PRL2018,Mojtaba,PhysRevLett.98.156601,PhysRevLett.101.036601,PhysRevLett.107.066604}. 
The frequency of the cavity mode is selected at $\omega_{c}/2\pi=10~\mathrm{GHz}$ with a cavity damping $\beta=1\times10^{-4}$ (corresponding to quality factor $Q \approx 5000)$. The reduced gyromagnetic ratio ($\gamma/2\pi$), damping-like torque efficiency ($\eta_{a}$), and field-like torque efficiency ($\eta_{b}$)  are taken as $2.8\times10^{6}~\mathrm{Hz/Oe}$, $0.2$, and $0.05$, respectively. For a Pt/FM bilayer, the typical range of damping-like torque efficiency ($\eta_{a}$) is 0.10 to 0.20 \cite{PhysRevLett.98.156601,PhysRevB.92.064426,doi:10.1021/acsaelm.1c01233} and the typical value of field-like torque efficiency ($\eta_{b}$) is $\approx$~0.05 \cite{PhysRevLett.101.036601,PhysRevLett.106.036601,PhysRevB.87.174417,Zhang2015}. Due to its low Gilbert damping parameter and high spin density, we choose yttrium iron garnet (YIG) as magnetic material. In particular, we consider a YIG film with a thickness $t=2\times10^{-5}~\mathrm{cm}$ (smallest thickness available commercially) and saturation magnetization, $M_s=144~\mathrm{emu/cm^3}$ \cite{Mojtaba}. For the calculation, the term $K_{F}K_{A}$ is taken as $5\times10^{-6}$ \cite{PhysRevLett.121.137203}. For YIG films, depending upon the thickness and preparation method, $\alpha$ varies from order $10^{-3}$ to $10^{-5}$ \cite{PhysRevLett.107.066604,doi:10.1063/1.3690918,doi:10.1063/1.4759039,doi:10.1063/1.4819157,PhysRevB.87.174417,PhysRevB.91.134407,PhysRevLett.111.106601,doi:10.1063/1.4852135,doi:10.1063/1.4861343,doi:10.1063/1.4896936}. Therefore, we vary $\alpha$ in our model from $3\times10^{-3}$ to $5\times10^{-5}$. 
Furthermore, we vary $J$ from $-5\times10^5~\mathrm{A/cm^2}$ to $5\times10^5~\mathrm{A/cm^2}$. The maximum value of chosen current density is at least one order of magnitude smaller than what is used for magnetic tunnel junctions (MTJs) \cite{Ikeda,Kang2021}. Note that, a current density of this order of magnitude has previously been reported for YIG/Pt systems \cite{PhysRevB.96.064407} to thermally  control  magnon-photon coupling in experiment. Reference \cite{PhysRevB.96.064407} reports that such current density leads to a rise of the system temperature above $40^{\circ}$C. Negative effects of heating on the magnetic properties can be drastically reduced by using a pulsed dc current through the Pt layer \cite{7836335} instead of using a continuous current. For instance, using a pulsed current with duty cycle of 50\%, heating effects can be mitigated while reaching reasonable high levels of current density between $-5\times10^6~\mathrm{A/cm^2}$ to $5\times10^6~\mathrm{A/cm^2}$ . {Such a high value of current density will create an Oersted field and, hence, modify the resonance condition. The generated Oersted field can be considered as a contribution to the effective magnetic field presented in Eq.~(2). Hence, this field will modify the resonance position of the magnon modes in the following way: for one polarity of the current density (J), the resonance field shifts up, while for the other, it shifts down. Experimentally, this affect can be compensated by tuning the biasing magnetic field so the resonance frequency remains the same.} In the following analysis, we consider two scenarios: (1) a relatively low continuous current density and (2) a higher pulsed current density. The effects of both conditions on the magnon-photon coupling process are compared below. The proposed experimental set up and measurement configuration is shown in Fig.~\ref{fig:schematic}.

\begin{figure*}[t]
\centering
\includegraphics[width = 1.0\textwidth, keepaspectratio]{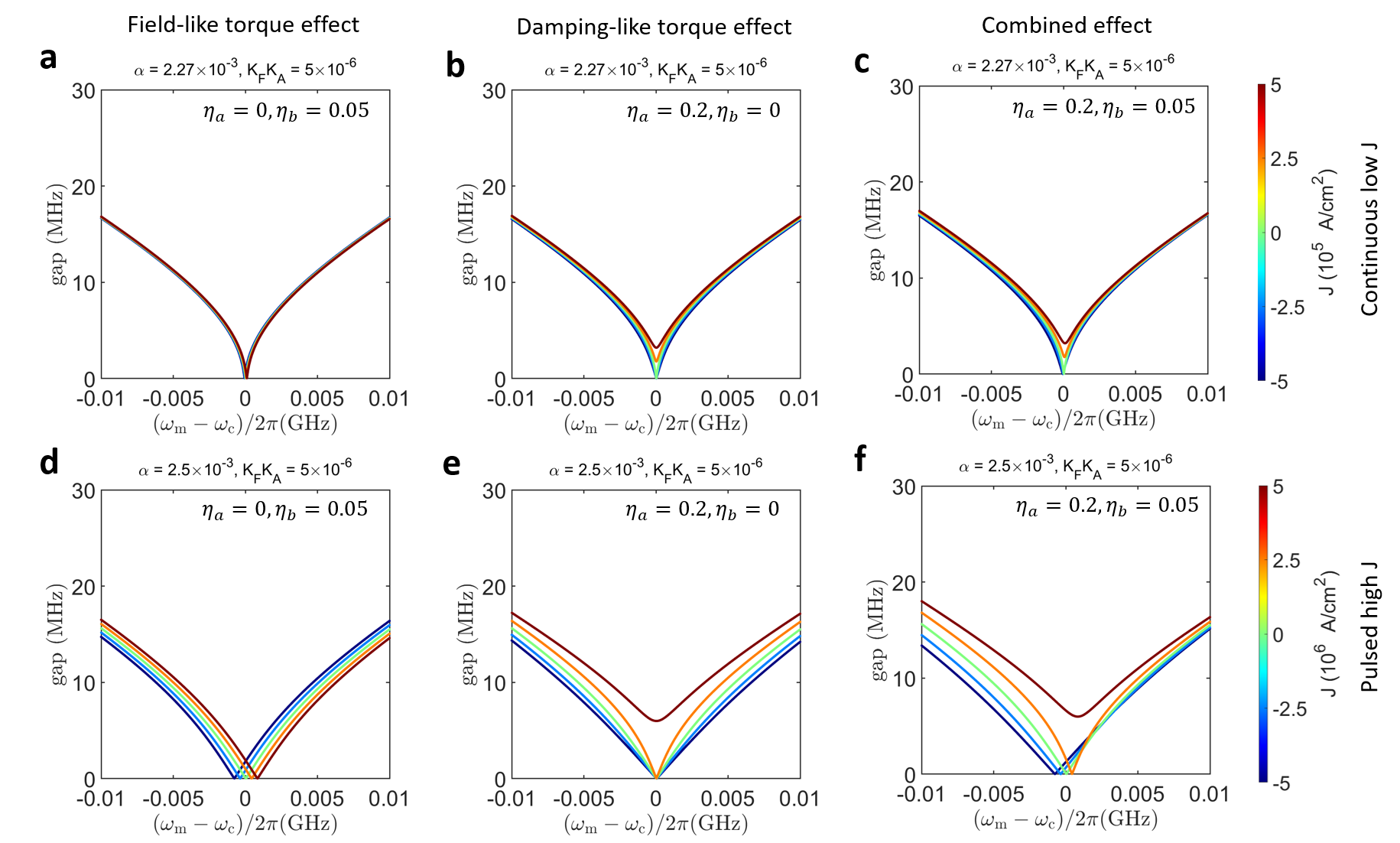}
\caption{Variation of the gap between two hybridized modes with respect to the field detunings ($\omega_{m}-\omega_{c}$) for  $\alpha=~2.27\times10^{-3}$ and continuous low $J$ from $-5\times10^5~\mathrm{A/cm^2}$ to $5\times10^5~\mathrm{A/cm^2}$ [(a),(b) and (c)] and for $\alpha=~2.5\times10^{-3}$ and pulsed high $J$ from $-5\times10^6~\mathrm{A/cm^2}$ to $5\times10^6~\mathrm{A/cm^2}$ [(d),(e) and (f)]. There is (a),(d) a horizontal shift in the location of the gap for $\eta_{a}=~0$ and $\eta_{b}~=0.05$ (pure field-like torque effect), (b), (e) vertical shift in the location of gap for $\eta_{a}=~0.2$ and $\eta_{b}=~0$ and (pure damping-like torque effect), and (c),(f) horizontal and vertical shifts for $\eta_{a}=~0.2$ and $\eta_{b}~=0.05$ (combined effect of damping-like and field-like torques).}
\label{fig:variation_gap_field_detuning}
\end{figure*}

\subsection{Dispersion and Linewidth}\label{section:dispersion and linewidth}
In Fig.~\ref{fig:Intermediate_alpha_low_J} (intermediate value of Gilbert damping parameter $\alpha$ and continuous, low value of current density $J$) and Fig.~\ref{fig:Intermediate_alpha_high_J} (intermediate value of $\alpha$ and pulsed, high value of $J$), the hybridized mode frequency ($\omega-\omega_{c}$) and linewidth ($\Delta{\omega}$) are plotted as a function of the field detuning ($\omega_{m}-\omega_{c}$). 

 We first focus on the former, shown in Fig.~\ref{fig:Intermediate_alpha_low_J}, top panels: For $\alpha = 2.27\times10^{-3}$ and for $J = 0~\mathrm{A/cm^2}$, {we observe a level attraction of the real part of the eigenvalues [Fig. 2 (b)] in a small region. For $J = -5\times10^5~\mathrm{A/cm^2}$, a similar behavior is found [Fig. 2(a)]}.
 However, the behavior drastically changes for reversed current polarity: for $J = 5\times10^5~\mathrm{(A/cm^2)}$, a gap {(a prominent level repulsion)} is seen between the hybridized modes. This clearly shows that depending upon the strength and direction of the dc current density $J$ 
one can tune the gap between the hybridized modes, i.e., transitioning the system into the strong coupling regime. Let us now consider Fig.~\ref{fig:Intermediate_alpha_high_J}, top panels: A similar but enhanced behavior can be observed for higher values of $J$ (i.e., $\vert J \vert=5\times 10^6$A/cm$^2$) and $\alpha = 2.5\times10^{-3}$ [Fig.~3]. As is obvious from Figs.~\ref{fig:Intermediate_alpha_low_J} and \ref{fig:Intermediate_alpha_high_J}, there is a shift in the position where the coherent coupling occurs. For negative and positive values of $J$, the resonance shifts towards the negative and positive sides of the field detuning ($\omega_{m}-\omega_{c}$), respectively. This shift can be understood by the fact that different magnitudes of field-like torques directly affect the resonance condition as will be discussed in Sec.~\ref{section:gap}. 

Next, we discuss the lower panels of Figs.~\ref{fig:Intermediate_alpha_low_J} and \ref{fig:Intermediate_alpha_high_J}. The linewidths of the two hybridized modes {distinctly} cross each other for $J = 5\times10^5~\mathrm{A/cm^2}$ and $J = 5\times10^6~\mathrm{A/cm^2}$ as is expected for {a broad coupling region} [Fig.~\ref{fig:Intermediate_alpha_low_J}(f) and Fig.~\ref{fig:Intermediate_alpha_high_J}(f)]. {This effect is less distinct f}or the cases $J = -5\times10^5~\mathrm{A/cm^2}$, $J = -5\times10^6~\mathrm{A/cm^2}$ and $J = 0~\mathrm{A/cm^2}$. {However, despite the lower number of region in the crossing regime (the crossing is less spread), we emphasize that level crossings} in the linewidths of the hybridized modes are {are also observed here}. {We note that the coupling region broadens as the current density increases from negative values to positive values [Figs.~\ref{fig:Intermediate_alpha_low_J}(d,e,f) and Figs.~\ref{fig:Intermediate_alpha_high_J} (d,e,f)]} and finally {a distinct crossing of linewidths is observed over a broad range [Fig.~\ref{fig:Intermediate_alpha_low_J}(f) and Fig.~\ref{fig:Intermediate_alpha_high_J} (f)]}.

{For special cases discussed in Sec. II of the SM, a level attraction [Fig. S1 (a,b)] in the real part and level repulsion [Fig. S1 (d,e)] in the imaginary part of the eigenvalues are observed along with exceptional points (EPs) \cite{Heiss_2004, doi:10.1126/sciadv.aax9144,PhysRevLett.123.237202,doi:10.1126/science.aar7709}. For more details on the observed EP we refer to the SM.}

\subsection{Anticrossing gap between hybridized modes}\label{section:gap}
Figure~\ref{fig:variation_alpha_gap_J} shows the variation of the anticrossing gap between the hybridized modes for different values of $\alpha$ and $J$. {It is clear that the variation is nonlinear in nature.} As is evident from the figure, the gap between the two hybridized modes becomes smaller for a larger value of $\alpha$. On the other hand, the gap also depends on the dc current density $J$: {the value of $\alpha$ for which the gap is {very small} increases if we go from from negative to positive value of the dc current density.} For a positive value of $J$, we also observe the gap between the hybridized modes slowly increases as $\alpha$ increases and becomes maximum for a particular value of $\alpha$, and then decreases if we further increase the value of $\alpha$, as shown in the inset of Fig~\ref{fig:variation_alpha_gap_J}(f). {For the low $\alpha$ regime, the anticrossing gap remains nearly the same for different orders of magnitude and directions of current density, as is shown in Figs. S2, S3 and S4 of the SM. However, for the high $\alpha$ regime, we observe a level repulsion in the real part and a level crossing in the imaginary part of the eigenvalues for different orders of magnitude and directions of the current density, as is shown in Figs. S5, S6 and S7 of the SM.}
For a different coupling strength ($K_FK_A$), we observe a similar trend. A positive current density is needed to increase the gap between the two hybridized modes as is shown in Fig. S8 (SM).

In the following discussion, we chose Gilbert damping parameters of $\alpha = 2.27 \times10^{-3}$ and $\alpha = 2.5 \times10^{-3}$ for different orders of magnitude of $J$ where {a very small anticrossing gap} for zero current density {is seen}, as illustrated in Fig.~\ref{fig:variation_alpha_gap_J}. For large $\alpha~(=~4\times10^{-3})$ and for very low $\alpha~(=~5\times10^{-5})$, the hybridized mode frequency ($\omega-\omega_{c}$) and the linewidth ($\Delta{\omega}$) plotted as a function of the field detuning ($\omega_{m}-\omega_{c}$) are shown in the Fig. S1 and Fig. S2 of the SM. Figure~\ref{fig:variation_gap_field_detuning} shows the variation of the magnitude of the gap between the two hybridized modes with respect to field detuning ($\omega_{m}-\omega$) for $\alpha=~2.27\times10^{-3}$ and $\alpha=~2.5\times10^{-3}$. For a pure field-like torque effect, there is a horizontal shift [as shown in Figs.~\ref{fig:variation_gap_field_detuning}(a) and \ref{fig:variation_gap_field_detuning}(d)] of the gap between the hybridized modes towards the positive value of field detuning as we go from negative to positive values of the current density $J$. However, for a damping-like torque effect, there is a vertical shift [as shown in Figs.~\ref{fig:variation_gap_field_detuning}(b) and \ref{fig:variation_gap_field_detuning}(e)] of the minimum gap (anticrossing gap): the anticrossing gap {increases} if we go from negative to positive values of $J$. For the combined field-like and damping-like torques effect, there are both horizontal and vertical shifts as can be seen in Figs.~\ref{fig:variation_gap_field_detuning}(c) and \ref{fig:variation_gap_field_detuning}(f). However, the horizontal shift due to the effect of field-like torque, vertical shift due to damping-like torque and the combined shift due to both field and damping-like torques are more pronounced for higher magnitudes of $J$ leading to an unusual behavior as is shown in panel (f).

Finally, we note that introducing the field-like and damping-like torques in the Landau-Lifshitz-Gilbert equation and coupling it with cavity mode through combined Amp\`ere and Faraday effects do not produce level attraction. The coupling term in our analysis is not affected by the $\delta$ term [see Eq. (\ref{main equation})], which is the parameter governed by spin torque. This means that transitioning the system from strong coupling to dissipative coupling and vice versa cannot be achieved by spin-transfer torques. 
\newline
\section{\label{sec:level4}Summary}
By coupling of the generalized LLG equation with the RLC equation of the cavity,  we revealed the coupling between magnon and photon modes under the influence of damping and field-like torques. Our results indicate that the magnitude of the level repulsion (manifested by the frequency gap of the hybridized modes) and, hence, the magnon-photon coupling strength can efficiently be controlled by varying the magnitude and the direction of dc current density $J$ for realistic parameters of the magnetic properties. Our model suggests that an on-demand manipulation of the magnon-photon coupling strength can be achieved for current densities of the order as small as $10^5$~A/cm$^2$ and an intermediate Gilbert damping of the order $10^{-3}$. Higher values of $J$ can definitely enhance the effect of damping and field-like torques on the magnon-photon coupling provided we use pulses of dc current to reduce possible heating effects. Therefore, the experimental realization of the proposed magnon-photon control mechanism should be feasible in YIG/Pt hybrid structures.

\section*{Acknowledgment}
{We acknowledge fruitful discussions with Dr. J. Sklenar (Wayne State University) and Dr. J. Q. Xiao (University of Delaware).}
Research supported by the U.S. Department of Energy, Office of Basic Energy Sciences, Division of Materials Sciences and Engineering under Award DE-SC0020308.

\bibliography{apssamp}

\end{document}